Far Infrared Edge Photoresponse and Persistent Edge Transport in an Inverted InAs/GaSb Heterostructure

G. C. Dyer[1], X. Shi[1,2], B. V. Olson[1], S. D. Hawkins[1], J. F. Klem[1], E. A. Shaner[1] and W. Pan[1]

[1]*Sandia National Laboratories, P.O. Box 5800, Albuquerque, New Mexico 87185*
[2]*Department of Physics, The University of Texas at Dallas, Richardson, Texas 75080*

Direct current (DC) transport and far infrared photoresponse were studied an InAs/GaSb double quantum well with an inverted band structure. The DC transport depends systematically upon the DC bias configuration and operating temperature. Surprisingly, it reveals robust edge conduction despite prevalent bulk transport in our device of macroscopic size. Under 180 GHz far infrared illumination at oblique incidence, we measured a strong photovoltaic response. We conclude that quantum spin Hall edge transport produces the observed transverse photovoltages. Overall, our experimental results support a hypothesis that the photoresponse arises from direct coupling of the incident radiation field to edge states.

InAs/GaSb double quantum well (DQW) structures with inverted type-II band alignment have attracted a great deal of current interest because they support the quantum spin Hall (QSH) effect,[1] in which these two-dimensional (2D) topological insulators (TI) display conductive edge channels and an insulating bulk state.[2,3] The QSH edge states are helical in nature, with each edge channel carrying a pair of spin-polarized, counter-propagating components that are topologically-protected from backscatter by time-reversal symmetry. Experimentally, research on InAs/GaSb devices thus far has focused primarily upon direct current (DC) transport phenomenology[2-10] at sub-Kelvin cryogenic temperatures in order to minimize residual bulk conductivity and accentuate QSH edge transport. There have been to this point few activities directed towards far infrared characterization the InAs/GaSb system.[11] On the other hand, it can be expected that photon induced redistribution of carriers can strongly affect the edge transport in the QSH effect, and may reveal optical techniques for manipulating spin-polarized carriers.

In this *Letter*, we present evidence of an incident far infrared field directly coupling to QSH edge states in an InAs/GaSb DQW structure, and develop a phenomenological description through examination of robust DC edge transport in the presence of dominant bulk conduction. The characterized InAs/GaSb Hall bar device, pictured in Fig. 1(a), has three gate terminals, denoted G1, G2 and G3, and eight Ohmic contacts, labeled C0 to C7 moving clockwise from the far left. Each of the Ohmic probes extending from the channel is 5 μm wide, with each probe separated from adjacent probes by 5 μm along the upper edge of the device. The 4 μm wide gates are situated, within alignment tolerance, centrally between contacts C1, C2, C3 and C4. From C0 to C5, the total length of the channel is 60 μm, with a width of 10 μm where Ohmic probes are absent.

The fabricated InAs/GaSb device is based upon a 14 nm InAs, 4 nm GaSb DQW structure[5,12] bookended by 50 nm AlSb layers with a 2 nm InAs cap. Assuming *a priori* the possibility of edge transport, the equivalent circuit representation in Fig. 1(b) includes non-identical upper and lower edge channels in parallel with bulk 2D conduction.[13] The electronic band structure of the studied InAs/GaSb DQW is plotted in Fig. 1(c), calculated using a 14-band ***K•p*** model.[14,15] This calculation clearly shows the hybridization gap between the electron ground state in the InAs quantum well and the heavy hole ground state in the GaSb quantum well and highlights appreciable spin-splitting in the bulk bands due to spin-orbit interaction (SOI). This material is thus a candidate to display robust intrinsic spin Hall edge transport when electrostatically doped into a bulk conducting state[16-19] in addition to supporting the QSH effect in the TI phase when the Fermi level is tuned to the hybridization gap.[1-10]

To first develop understanding of the device transport behavior, we consider a set of complementary DC measurements in Fig. 2. A sinusoidal 11 Hz current $I_{50} = 500$ nA was applied between contacts C5 and C0 to measure the four-terminal resistances $R_{50,ij} \equiv (V_i - V_j)/I_{50}$ using standard lock-in techniques. Gate G2 had an applied DC bias $V_{G2}$ while the other two gates, as well as C0, remained fixed at ground potential. Although all three gates tuned the device transport self-consistently, we focus here on G2 because it represents a mirror symmetry line between C0 and C5 and isolates built-in differences in the upper and lower edges of the device.

The four-terminal resistances measured at T = 8 K in Fig. 2(a) reflect the presence of a charge neutrality point (CNP) near $V_{G2}$ = -2.8 V. Surprisingly, transverse resistances $R_{50,17}$ and $R_{50,46}$ that are opposite in polarity but nearly identical in magnitude exist for all applied $V_{G2}$ biases. The fact that all four-point resistances are much smaller than $h/e^2$ indicates that bulk conductance dominates over edge transport. However, the relation $-R_{50,17} = R_{50,46}$ cannot be explained by a bulk conduction mechanism. This shows that that even in our device of macroscopic size and at an elevated temperature of 8 K edge conductance still plays an important role in four-terminal resistance values. Given the prominent SOI inherent to the band structure in Fig. 1(c), edge transport in both the bulk conducting and TI phases is consistent with the measured transverse resistances. The longitudinal resistances $R_{50,32}$, $R_{50,41}$ and $R_{50,67}$, meanwhile, follow the relations $3R_{50,32} < R_{50,41}$ and $R_{50,41} < R_{50,67}$. In the context of Fig. 1(b), it follows that $3R_{50,32} \sim R_{50,41}$ because $R_{U1} \cong R_{U2} \cong R_{U3}$. $R_{50,32}$ also tunes with $V_{G1}$ and $V_{G3}$ (not plotted), a characteristic in agreement with the circuit in Fig. 1(b) and again inconsistent with purely bulk transport.

Consideration of the temperature dependence of the four-terminal DC resistances at $V_{G2}$ = -2.8 V in Fig. 2(b) aids in elucidating the underlying phenomenology. The longitudinal resistances $R_{50,67}$ and $R_{50,41}$ systematically decrease with temperature while both transverse resistances $R_{50,17}$ and $R_{50,46}$ increase in magnitude with temperature to around 80 K. First, this cannot be attributed to bulk transport alone because any measured transverse resistance $R_{xy}$ due to a Hall probe misalignment of displacement $\delta L$ is proportional to the bulk resistivity $\rho_{bulk}$, $R_{xy} \approx \rho_{bulk} \delta L / W$, where $W$ is the channel width. Assuming the circuit in Fig. 1(b), an equivalent two-terminal resistance $R_{eq}$ can be defined as $R_{eq}^{-1} = R_{50,50}^{-1} = R_{bulk}^{-1} + R_{top}^{-1} + R_{bot}^{-1}$, where $R_{top} = 2R_A + R_U$ and $R_{bot} = 2R_A + R_L$ are the series resistances along upper and lower device edges, respectively, and $R_U = R_{U1} + R_{U2} + R_{U3}$. Thus, the four-terminal resistances in terms of lumped resistive elements are $R_{50,67} = R_{eq} R_L / R_{bot}$, $R_{50,41} = R_{eq} R_U / R_{top}$, $R_{50,17} = R_{eq} R_A (R_L - R_U) / R_{top} R_{bot}$, and $R_{50,46} = R_{eq} R_A (R_U - R_L) / R_{top} R_{bot}$. Each edge resistive element ($R_A$, $R_{U1}$, $R_{U2}$, $R_{U3}$, $R_L$) if understood as a one-dimensional quantum channel without spin degeneracy has a resistance of at least $h/e^2 \sim 25.8$ k$\Omega$.[6] Furthermore, the measured longitudinal resistances indicate that $R_{bulk}$ is on the order of several hundred Ohms. This implies that $R_{eq} \cong R_{bulk}$, and measured longitudinal resistances are therefore approximately proportional to the bulk resistance.

From the expressions for the transverse resistances, it is apparent that transport measurements require $R_L > R_U$, in contrast with a ballistic Landauer-Büttiker QSH transport description[20] in the absence of bulk conduction where $R_{50,41} > R_{50,67}$ and $R_L < R_U$. The phase-breaking probes C2 and C3 effectively short circuit a portion of the upper edge, resulting in three 5 μm channels in series with $R_U \geq 3\, h/e^2$. However, incoherent QSH transport where the phase-breaking mean free path is significantly shorter than the 25 μm length[21] along the lower edge channel can nonetheless produce the observed transport behavior. Thus, we conclude that the measured transport characteristics result from incoherent transport along at least the lower edge channel, as illustrated in the magnified portion of Fig. 1(b) as a series of short helical QSH edge channels with broken phase coherence.[22-24]

To characterize the far infrared photoresponse of the device, we measured the photovoltages $\delta V_{ij} \equiv (V_i - V_j)$ with 180 GHz radiation normally incident on the sample through z-cut quartz cryostat windows. A set of Virginia Diodes Inc. Schottky multipliers driven by an RF local oscillator and modulated at 75 Hz with 50% duty cycle provided a peak power of $P_0 = 3.9$ mW. We applied no DC current ($I_{50} = 0$) and tuned G2 because it presents a mirror symmetry line with respect to the device bulk. Furthermore, the sample underwent alignment to reduce signal artifacts resulting from spatial inhomogeneity[25] in the incident far infrared intensity. Because the respective equivalent circuits linking C6 to C1 and C4 to C7 are identical when G2 is tuned, $(V_6 - V_1) = (V_4 - V_7)$ under any DC bias or optical excitation symmetric about G2. The nearly identical diagonal photovoltages $\delta V_{61}$ and $\delta V_{47}$ shown in Fig. 3(a) thus indicate any spurious Seebeck or rectification signals were virtually eliminated via this optical alignment.

The longitudinal photovoltages $\delta V_{67}$ and $\delta V_{41}$ and transverse photovoltages $\delta V_{46}$ and $\delta V_{17}$ measured at 8 K in Figs. 3(a) and 3(b), respectively, bear several similarities to the analogous DC measurements in Fig. 2(a). Most noticeably in the vicinity of the CNP, $\delta V_{67} > \delta V_{41}$ and $\delta V_{46} = -\delta V_{17}$, implying that the photoresponse is correlated with asymmetries between upper and lower edge channels. Furthermore, in Fig. 2(b) it is evident that $R_{50,46} \neq -R_{50,17}$ and $R_{50,61} \neq R_{50,47}$ beginning in the 40-50 K temperature range, possibly indicating a breakdown in edge transport. The photovoltage measured as a function of temperature in Figs. 3(b) and 3(c) is negligible above 45 K, again suggesting a link between the photovoltaic signal and edge conduction.

However, there are also several critical discrepancies between the photoresponse and DC transport measurements. The longitudinal and transverse photovoltages in Figs. 3(a) and 3(b) are all on the order several µV, in contrast with the four-terminal resistances that span an order or magnitude in Fig. 2(a). Furthermore, the photoresponse $\delta V_{67}$ in Figs. 3(c) and 3(d) decreases precipitously with temperature at all gate biases, also in contrast with the more gentle temperature dependence of the four-terminal DC resistances in Fig. 2(b). These conspicuous differences suggest that the measured photovoltages are not driven by a bulk response, otherwise the far infrared photoresponse would largely mirror the DC transport at any fixed operating point.

Consideration of two common bulk 2D response mechanisms, semi-classical 2D plasmonic[26] homodyne mixing[27-29] and a photo-thermoelectric response,[30-33] supports this premise. Both response mechanisms have a strong dependence on the spatial location of any experimental asymmetry and track the DC transport as $\sigma^{-1} \partial\sigma/\partial V_{G2}$,[28,30] where $\sigma$ is the bulk channel conductivity. Because $R_{bulk} \ll R_{bot}$, we may approximate $\sigma \propto 1/R_{50,67}$ and $\partial\sigma/\partial V_{G2} \propto \partial(1/R_{50,67})/\partial V_{G2}$, letting $G = 1/R_{50,67}$ to define $\chi \equiv G^{-1} \partial G/\partial V_{G2}$. This expression is plotted in Fig. 3(e) as a function of both gate voltage and temperature. In clear contrast with the photoresponse in Figs. 3(a)-(d) that drops rapidly with temperature and peaks at the CNP, $\chi$ has weak temperature dependence, approaches zero at the CNP, and changes polarity when majority carriers below G2 shift from electrons to holes. Given a characterization methodology that minimized experimental asymmetries, it is unsurprising that there is no evidence to support a bulk photoresponse predicated on asymmetry.

The photovoltaic response under tuning of G2 is further explored through polarization-dependent measurements in Fig. 4. A pair of wire grid polarizers, one fixed to project half of the total incident power along $\theta_0 = 135°$ and the other freely rotating to project a fraction of this power along $\theta_{pol}$, enabled characterization of the polarization dependent response. In Fig. 4(a), the 180 GHz relative electric field amplitudes and orientations relative to Fig. 1(a) are illustrated. The responsivity is defined as $r_{ij} \equiv \pi\sqrt{2}\delta V_{ij}/2P(\theta_{pol})$, where $P(\theta_{pol}) = P_0 cos^2(\theta_{pol} - \theta_0)/2$. This normalizes the response by accounting for the peak-to-peak signal generated from the square wave modulated far infrared source, but neglects possible corrections for sample area and window loss. Figs. 4(b) and 4(c) show the longitudinal responsivity $r_{67}$ under rotation of $\theta_{pol}$ with fixed $V_{G2} = -2.8$ V and with $V_{G2}$ tuned, respectively. In the vicinity of the CNP, the response is strongest with incident field oriented longitudinally along $\theta_{pol} = 0°$ or 180° and there is little response with $\theta_{pol} = 90°$, within uncertainties in device and polarizer orientations.

The excitation frequency of 180 GHz corresponds to a 0.7 meV photon energy, comparable to both the hybridization gap $\Delta$ (~ 1-4 meV)[2,5,12] and $kT$ at 10 K (~ 0.9 meV). In light of the strong temperature dependence of the photovoltage, this suggests that hybridization physics may play a role in the response. Given that $h\nu < \Delta$, virtual photoconductivity[34] driven by the incident alternating current (AC) field along $\theta_{pol} = 0$ could contribute to a bulk rectified current. The lack of experimental asymmetry permits no net time averaged virtual photocurrent, though, and argues against this type of bulk response.

Instead, we posit that direct coupling of the AC radiation field to edge modes below G2 produces the photoresponse. Two possible mechanisms include the generation of a DC photocurrent[35] in non-ballistic, edge channels and the rectification of QSH edge plasmons.[36] The latter mechanism would be accompanied by spin rectification of the spin-polarized plasma density fluctuations.[37] Because of the broken translational symmetry along non-identical upper and lower edges, linearly polarized radiation along $\theta_{pol} = 0$ is appropriate to both couple with the device edge channels and produce a non-zero net response.

The longitudinal signals $\delta V_{67}$ and $\delta V_{41}$ observed in Fig. 3(a) arise because the two components of the spin-polarized helical currents propagate away from the edges below G2 in opposing directions. Whether viewed as single particle DC currents or rectification of collective AC currents, the spin up and down components have respective induced DC current densities $\delta J_\uparrow = \pm \delta J$ and $\delta J_\downarrow = \mp \delta J$ that traverse symmetric impedances $Z$ such that $V_\uparrow = -V_\downarrow$ at opposing sides of G2. The chemical potential must shift by $+V_\uparrow$ on one side of G2 and by $-V_\uparrow$ on the other side, thus conserving charge along both the upper and lower device edges. This also may explain the temperature dependence of the photoresponse since reduction in the QSH mean free path should degrade the induced coherent current $\delta J$.

Letting $V_4 = +\frac{1}{2}V_U$, $V_1 = -\frac{1}{2}V_U$, $V_6 = +\frac{1}{2}V_L$ and $V_7 = -\frac{1}{2}V_L$, where U and L denote upper and lower device edges, respectively, facilitates comparison of the experiment to the above description. This results in $\delta V_{41} = V_U$, $\delta V_{67} = V_L$, $\delta V_{46} = \frac{1}{2}(V_U - V_L)$, $\delta V_{17} = \frac{1}{2}(V_L - V_U)$, and $\delta V_{61} = \delta V_{47} = \frac{1}{2}(V_U + V_L)$, for the respective longitudinal, transverse and diagonal photovoltaic responses. Strikingly, $\delta V_{46} = -\delta V_{17}$ and $\delta V_{61} = \delta V_{47}$ in Figs. 3(a) and 3(b), fully consistent with this heuristic explanation as well as prior discussion of the DC transport.

In summary, we have observed a far infrared photoresponse consistent with direct AC driving of QSH edge currents in an InAs/GaSb DQW field effect device that supports a TI phase. Additionally, DC transport measurements have demonstrated that edge conductance remains important in our device of macroscopic size, even in the presence of significant bulk conduction and outside of the ballistic transport limit. QSH conductance, both in the TI and bulk conductive states, offers a potential explanation for the combined transport and photoresponse phenomenology. Our results point towards an open and potentially rich path of inquiry analogous to work already begun on optical[25] and far infrared[38] surface excitation in 3D TIs.

This work was supported by the Department of Energy, Office of Basic Energy Science, Division of Materials Sciences and Engineering. Sandia National Laboratories is a multi-program laboratory managed and operated by Sandia Corporation, a wholly owned subsidiary of Lockheed Martin Corporation, for the U.S. Department of Energy's National Nuclear Security Administration under contract DE-AC04-94AL85000. The authors thank Michael Flatté at the University of Iowa for use of the *K•p* software that contributed to this work.

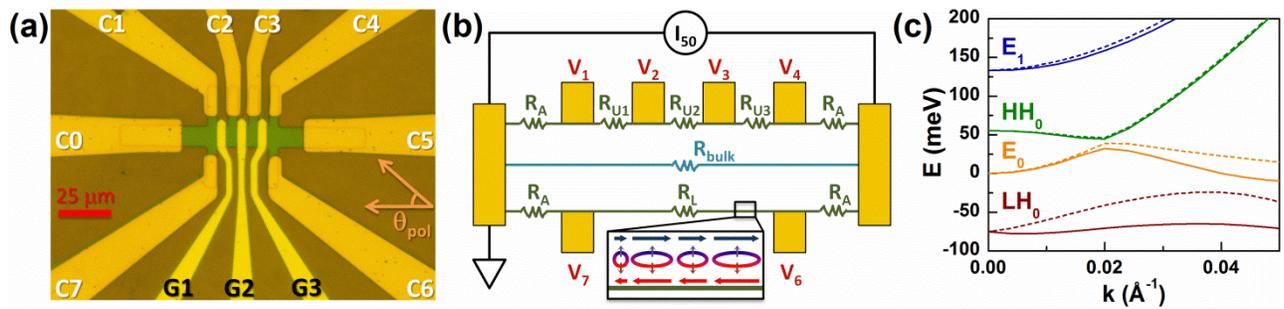

**Figure 1**. (a) Micrograph of a 10 μm wide by 60 μm long InAs/GaSb Hall bar device with three independently biased gates and eight Ohmic terminals. (b) A circuit diagram illustrating the experimental bias scheme and the device's equivalent resistive network. The magnified region conceptually shows incoherent edge transport where scattering centers repeatedly break phase coherence over the length of the channel. (c) Band structure of the InAs/GaSb DQW material system showing hybridization of the electronic ground state (orange) with lowest heavy hole subband (green). Additional electron (blue) and light hole (red) subbands are shown, as well as the spin-splitting of each band (solid and dashed lines).

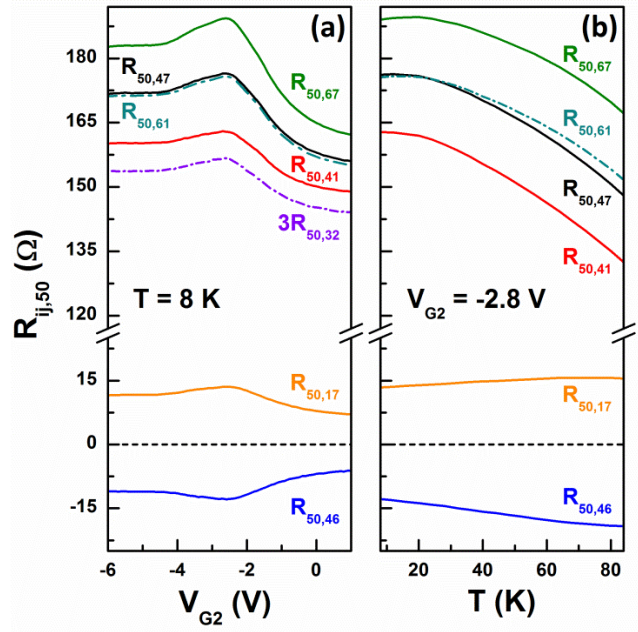

**Figure 2**. The longitudinal resistances $R_{50,67}$ (green) and $R_{50,41}$ (red), the transverse resistances $R_{50,46}$ (blue) and $R_{50,17}$ (orange), and the diagonal resistances $R_{50,47}$ (black) and $R_{50,61}$ (broken teal) are shown as a function of (a) gate voltage $V_{G2}$ at T = 8 K and (b) temperature at $V_{G2}$ = -2.8 V. Additionally, the longitudinal resistance $R_{50,32}$ (broken violet) is plotted in (a) as $3R_{50,32}$.

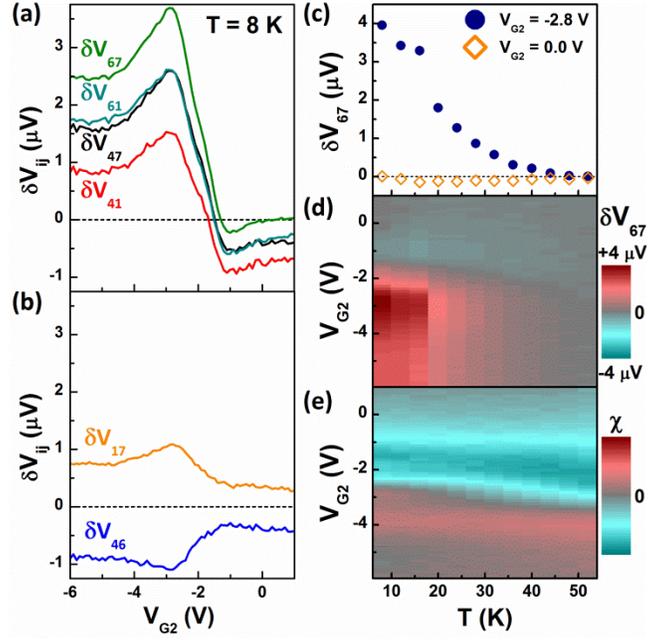

**Figure 3**. (a) The longitudinal photovoltages $\delta V_{67}$ (green) and $\delta V_{41}$ (red) and the diagonal photovoltages $\delta V_{47}$ (black) and $\delta V_{61}$ (teal) are plotted as a function of gate voltage $V_{G2}$ at T = 8 K. (b) The transverse photovoltages $\delta V_{46}$ (blue) and $\delta V_{17}$ (orange) are plotted as a function of gate voltage $V_{G2}$ at T = 8 K. (c) $\delta V_{67}$ is plotted as a function of temperature for $V_{G2} = -2.8$ V (blue circles) and $V_{G2} = 0.0$ V (orange diamonds). (d) $\delta V_{67}$ is mapped as a function of both temperature and gate voltage $V_{G2}$. For all photoresponse measurements in (a)-(d), normally incident 180 GHz radiation was polarized along the length of the channel. (e) The approximate bulk transport characteristic $\chi = G^{-1} \partial G / \partial V_{G2}$, where $G = 1/R_{50,67}$, is mapped as a function of temperature and gate voltage $V_{G2}$.

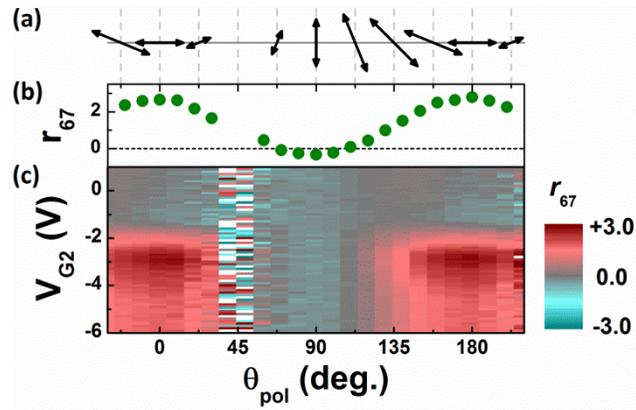

**Figure 4**. (a) The relative amplitude and polarization of the incident 180 GHz field is shown as a function of the rotatable polarizer orientation $\theta_{pol}$. (b) The responsivity $r_{67}$ (green circles) at T = 8 K with 180 GHz radiation incident is shown as a function of $\theta_{pol}$ with $V_{G2}$ = -2.8 V. (c) The responsivity $r_{67}$ at T = 8 K with 180 GHz radiation incident is mapped as a function of $\theta_{pol}$ and gate voltage $V_{G2}$. In both (b) and (c), the responsivity $r_{67}$ has units mV/W and missing data correlates with $\theta_{pol} = 45°$ where the incident power was negligible.

13